\documentclass[tightenlines,aps,prc,showpacs,amsmath,showkeys,
superscriptaddress]{revtex4}
\usepackage[dvips]{graphicx}
\usepackage{overpic}

\begin{document}

\title{\textbf{Future directions in kaonic atom physics}}

\author{E.~Friedman}
\email{elifried@cc.huji.ac.il}
\affiliation{Racah Institute of Physics, The Hebrew University,
Jerusalem 91904, Israel}

\date{\today}

\begin{abstract}
Recent progress and open problems in kaonic atom
physics are presented. A
connection between phenomenological deep potentials and the underlying
$K^-N$ interaction is established as well as the need for a theory for
multinucleon absorption of kaons. $K^-$ absorption at rest 
to specific $\Lambda $ hypernuclei states is briefly
discussed.
\end{abstract}

\pacs{13.75.Jz, 25.80.Nv, 36.10.-k}

\keywords{$K^-N$ interaction,kaonic atoms}

\maketitle
\section{Introduction and Background} 
\label{intro}
Attempting to foresee directions of further research into
the field of kaonic atoms one naturally focuses on open problems
on the one hand and on recent progress on the other. 
All the data on strong interaction effects in medium-weight and
heavy kaonic atoms come from experiments in the 1970's 
\cite{FGa07}, using solid targets and a single GeLi X-ray detector.
Analyses of the data show good consistency between the various experiments,
and sub-sets of the data do not lead, in general, to conclusions
which differ from what is found from analyses of the whole set of data.
A new generation of experiments characterized by the use of many 
detectors and of tracking facilities for reducing 
background \cite{Iwa97}
and with excellent timing capabilities \cite{Baz11} 
addressed `puzzles' with
kaonic atoms of hydrogen and $^4$He.

Starting with medium-weight and heavy elements,
a long
standing problem with these kaonic atoms has been the question of the
depth of the real optical potential.
Deep potentials, in the range of ${\rm Re}\:V_{K^-}
(\rho_0)\sim -$(150-200) MeV are obtained in comprehensive global fits to
$K^-$-atom strong-interaction shift and width data by introducing empirical
density dependent (DD) effective $K^-N$ amplitudes \cite{FGB94}. In contrast,
considerably shallower potentials, in the range of ${\rm Re}\:V_{K^-}
(\rho_0)\sim -$(40-60) MeV, are obtained for zero kinetic-energy kaons 
when {\it threshold} chiral
scattering amplitudes are used and 
self energy (SE) contributions are included
in the in-medium corrections \cite{ROs00}. 
 The large difference between the two
results has been known as the `deep {\it vs.} shallow puzzle' 
and is of interest
not only because of the credibility of the two approaches but also because the
depth of the potential may have far-reaching consequences for the possible
existence of kaon-nucleon clusters. Very recently \cite{Cie11,Cie11a}
this problem was solved by using in-medium sub-threshold $K^-N$ scattering
amplitudes which reproduced the empirical 
characterization of deep potentials \cite{FGB94}, namely, that the
real potentials are compressed relative to the nuclear charge distribution.
These results demonstrated the need to supplement the theory with multi-nucleon
absorption terms.

For the very light kaonic atoms new experiments \cite{Iwa97,Baz11} 
removed the long-standing puzzle of kaonic hydrogen where now the results
are fully consistent with the rest of the available data on antikaon-nucleon
interactions near threshold. For $^4$He the new results 
\cite{Oka07} removed the conflict
of more than an order of magnitude between any credible theory and experiment.

The future directions outlined below result from this background.

\section{Sub-threshold amplitudes}
\label{sec:1}

\begin{figure*}
  \includegraphics[width=0.75\textwidth]{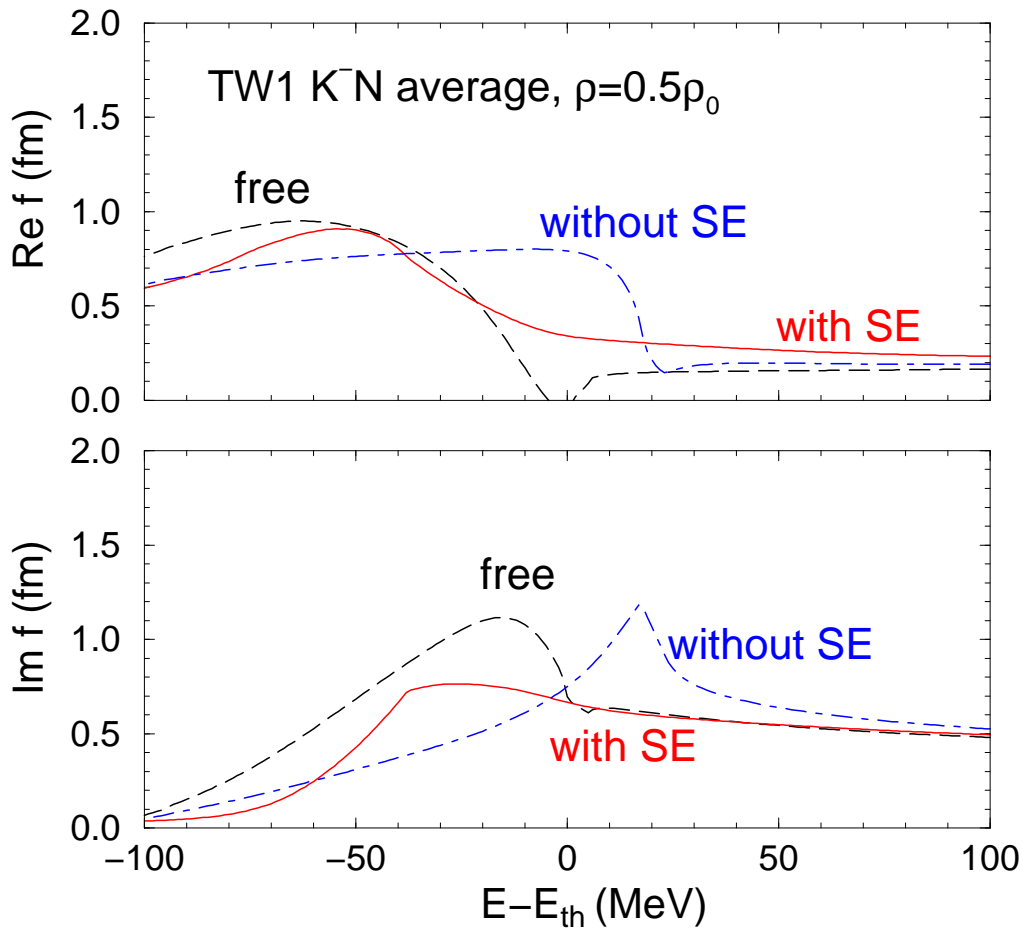}
\caption{ Examples for average scattering amplitudes for a
leading-order Tomozawa-Weinberg formulation.
Dashed curves: free space, dot-dashed: Pauli blocked amplitude (without SE)
at $\rho=0.5\rho_0$, solid curves: including meson and baryon self energies
(with SE) at $\rho=0.5\rho_0$.}
\label{fig:1}       
\end{figure*}

Introducing empirical density dependent scattering amplitudes into
global optical-model fits to data over the whole of the periodic table,
deep real potentials were obtained that produced superior fits compared
to fixed-amplitude approaches \cite{FGB94}. Another feature of these fits
was  a `compression' of the real
potential relative to the nuclear charge distribution.
This result means
that an underlying antikaon-nucleon amplitude must increase with the nuclear
density such that it overshadows any finite-range effect that might be 
present. This is a general result, independent of details. We therefore
focus attention on the density dependence of the $K^-N$ scattering amplitude,
where its energy dependence near threshold is strongly affected
by the proximity of the $\Lambda 1405$ resonance.  
Figure  \ref{fig:1} shows 
chirally leading-order Tomozawa-Weinberg $K^-N$ scattering amplitudes marked
TW1 solution, in free space and in symmetric nuclear medium for
half the nuclear density, where medium corrections with self-energy and
without are included, see refs.\cite{Cie11,Cie11a} for details.

%

The in-medium amplitudes are defined by the Mandelstam parameter  
$s=(E_{K^-}+E_N)^2-(\vec p _{K^-}^{~cm}+\vec p_N^{~cm})^2$
where the momenta are those implied by a kaon  bound in the atom
and by a nucleon which is part of the nucleus. Averaging over directions, 
using the
Fermi gas model for the nucleus and taking the local kinetic energy for
the kaon, one ends up with 
\begin{equation}
\sqrt s \approx E_{th} - B_N - \xi_NB_K -15.1(\rho/\rho_0)^{2/3}
 +\xi_K({\rm Re}~V_{\rm opt}+V_c).~({\rm MeV})
\end{equation}
in obvious notation, with $B_N$ an average binding energy for a nucleon 
and $\xi_N = m_N/(m_N+m_K), ~~~\xi_K =m_K/(m_N+m_K),~~~  E_{th}=m_K+m_N$.
We note that the potential is proportional to the amplitude, but the amplitude
depends on the potential through the argument $\sqrt s$. Therefore a self
consistent iterative procedure was applied at every radial point
for each nucleus in the data. That transforms the energy dependence into
density dependence.

\begin{figure*}
  \includegraphics[width=0.75\textwidth]{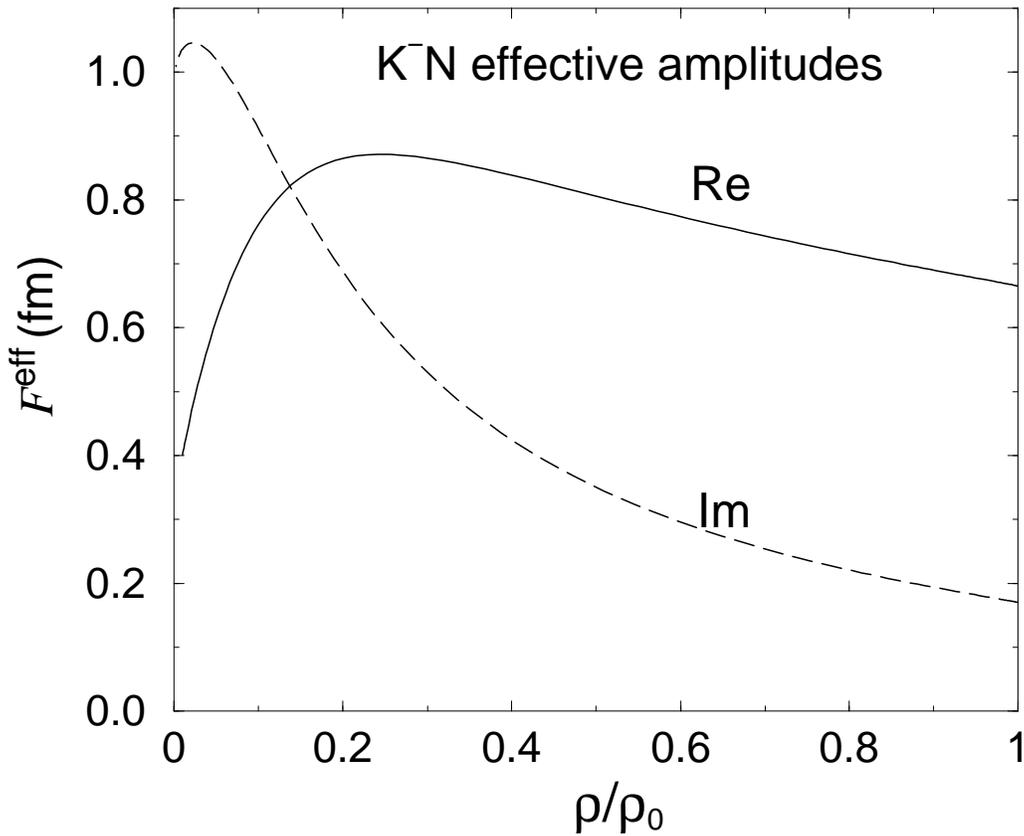}
\caption{Effective  amplitudes for $K^-$ Ni interaction.}
\label{fig:2}       
\end{figure*}

Figure \ref{fig:2} shows an example for the density dependence of the
scattering amplitude obtained for
the $K^-N$ interaction in Ni. 
Near the nuclear surface  the real part increases with density
whereas the imaginary part decreases,
in line with the compression and inflation of the real and imaginary 
potentials, respectively, observed in ref.\cite{FGB94}. The small values
of the imaginary part at large densities is due to the sharp drop
when energies go well below threshold, fig.\ref {fig:1}, 
towards the $\pi \Sigma$ threshold at $E-E_{th}\simeq -100$ MeV.

\section{Kaonic atom potentials}
\label{sec:2}
\subsection{Medium-weight and heavy nuclei}
\label{sec:21}
Comparisons between data and predictions made with optical potentials 
based on the above model lead to inadequate agreement with experiment,
with $\chi ^2$ per point of 10. Adding a phenomenological potential with
four adjustable parameters (for 65 data points) leads to $\chi ^2$ per point 
of 2. This phenomenological term is dominated by  $\rho ^2$ dependence,
suggesting indeed two-nucleon absorption which is not
included in the present $K^-N$ amplitudes. A real dispersive
term is to be expected.

\begin{figure*}
  \includegraphics[width=0.75\textwidth]{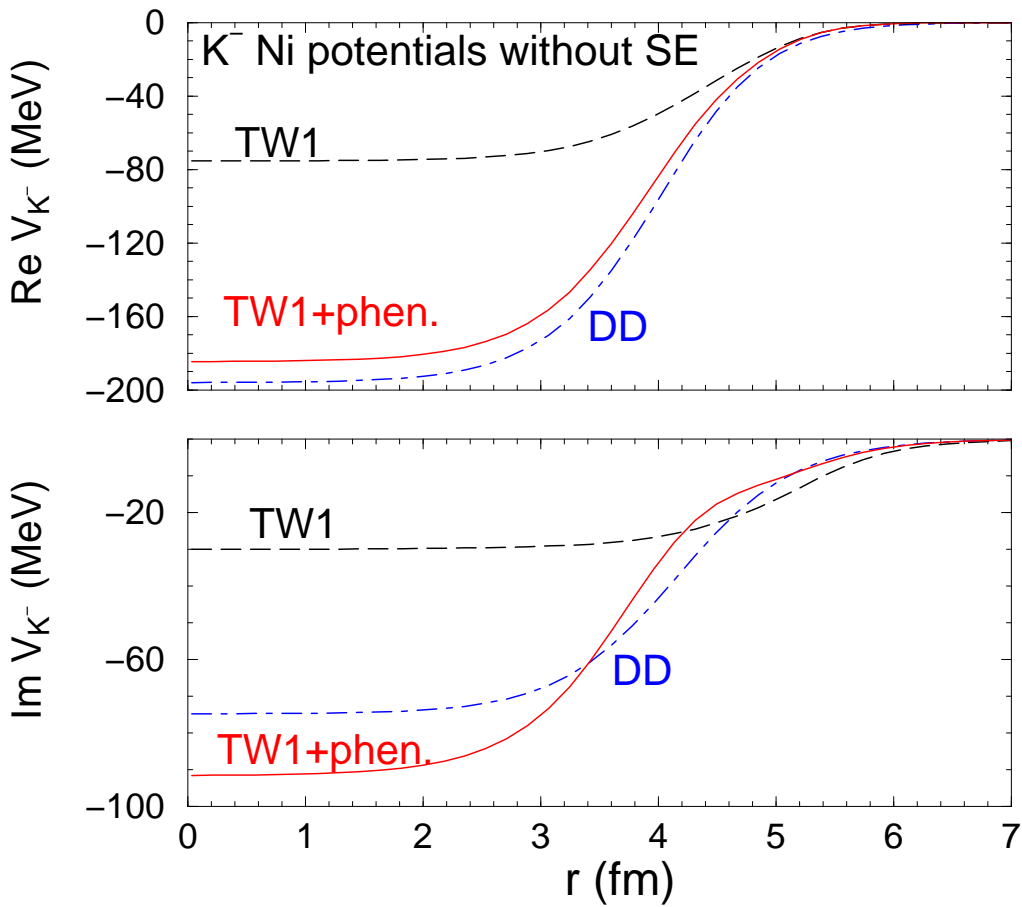}
\caption{ $K^-$ nuclear potentials for $K^-$ atoms of Ni.}
\label{fig:3}       
\end{figure*}

Figure \ref{fig:3} shows examples for the $K^-$Ni  potential. The real 
potential based on the TW1 amplitudes 
\cite{Cie11a} is 80 MeV deep compared to 40-50 MeV
deep in ref.\cite{ROs00}. The cause for the latter 
is that ref.\cite{ROs00} used the
amplitudes at threshold whereas 
figure \ref{fig:1} shows that at threshold the
real part of the SE version is $\approx$ 50\% of typical subthreshold values. 
For the full nuclear density the corresponding energy is as low as 
50 MeV below threshold. Also shown is  
the best fit potential when a phenomenological term is added,
and for reference, also the purely empirical DD potential. 
It is tempting to reject the
phenomenological additional term as being unacceptably large. However,
the values at the nuclear center are not the meaningful quantities 
as kaonic atoms are sensitive mostly to the potential close to the nuclear
surface \cite{FGa07}. Considering that strong interaction
effects in kaonic atoms are dominated by absorption, we note that the shape
of the imaginary potential in the surface region ($\approx$ 4 fm for Ni) is
modified by the phenomenological term
by $\pm$30\%, close to estimates of multi-nucleon absorptions
obtained from old emulsion and bubble-chamber experiments \cite{VVW77}. 
Replacing the additional 
phenomenological term by a multinucleon extension of the present
`microscopic' approach is a high priority line of research.

\subsection{Light nuclei}
\label{sec:22}
Formation rates of hypernuclear states following the 
$^AZ(K^-_{stop},\pi ^-)^A_\Lambda Z$ reaction have been another probe
for studying the interaction of $K^-$ mesons with nuclei at zero kinetic
energy, where experimental results have been limited mostly to light 
$p$-shell nuclei. 
The analysis of these reactions is the other extreme compared to
global analysis of kaonic atoms, by dealing separately not only with
each nuclear species but also by handling explicitly specific kaonic
atom states.
Recent analysis of experimental results from 
FINUDA \cite{Agn11} showed \cite{Cie11b} that with the sub-threshold
approach the deep $K^-$-nucleus potential is favored. This conclusion
is based on {\it relative} formation rates but the absolute scale
of these is not reproduced by the calculations. 
Understanding the absolute scale is another direction of work
in the field of kaonic atom physics.

Finally we turn to kaonic atoms of He. Table \ref{tab:1} summarizes
predictions made for $^{3,4}$He with a global optical potential 
 based on the TW1 amplitude plus a phenomenological
term. Other variations such as the removal of the phenomenological term
or the inclusion of a $p$-wave term change these values by up to 0.3 eV.
However, future high precision experiment, well beyond the present
capabilities, could bring interesting results.

\section{Conclusions}
\label{sec:con}

Having established a connection between {\it deep} real optical
potentials and underlying chiral-motivated $K^-N$ scattering amplitudes,
the next step will be to implement
a `microscopic' multinucleon absorption terms to supplement the one-nucleon
sub-threshold approach to kaonic atom potentials. 
Another near-future direction could be studies of the
absolute scale of formation rates in stopped kaon absorption experiments
where the current predictions underestimate experimental results by as much
as a factor six.  
Refinements of global analyses by studies of different
levels in a given kaonic atoms could bring 
interesting results. 
We note also that the role of a $p$-wave term in kaonic atoms has
not been established \cite{Cie11a}.
Last but not least, any new high quality
experiment will be most welcome.

%
%
\begin{table}
\caption{Shifts and widths for the 2$p$ state in $^{3,4}$He.}
\label{tab:1}       
\begin{tabular}{lll}
\hline\noalign{\smallskip}
   & $^3$He & $^4$He \\
 shift (eV) & 0.3 & $-$0.2 \\
 width (eV) & 2.1 & 1.6 \\
\noalign{\smallskip}\hline
\end{tabular}
\end{table}

\begin{acknowledgements}
I wish to thank  Ale\v{s} Ciepl\'{y}, Avraham Gal, Daniel Gazda and
Ji\v{r}\'i~Mare\v{s} who contributed much to the emerging sub-threshold
approach.
\end{acknowledgements}



\end{document}